\documentclass[twocolumn,showpacs,preprintnumbers]{revtex4}%
\usepackage{amssymb}
\usepackage{amsfonts}
\usepackage{amsmath}
\usepackage{graphicx}
\usepackage{dcolumn}
\usepackage{bm}%
\begin{document}
\title{Intermodulation Gain in Nonlinear NbN Superconducting Microwave Resonators}
\author{B. Abdo}
\email{baleegh@tx.technion.ac.il}
\author{E. Segev}
\author{O. Shtempluck}
\author{E. Buks}
\affiliation{Microelectronics Research Center, Department of Electrical Engineering,
Technion, Haifa 32000, Israel}
\date{\today}

\begin{abstract}
We report the measurement of intermodulation gain in NbN superconducting
stripline resonators. In the intermodulation measurements we inject two
unequal tones into the oscillator, the pump and signal, both lying within the
resonance band. At the onset of instability of the reflected pump we obtain a
simultaneous gain of both the idler and the reflected signal. The measured
gain in both cases can be as high as 15 dB, whereas to the best of our
knowledge intermodulation gain greater than unity in superconducting
resonators has not been reported before in the scientific literature.

\end{abstract}
\maketitle




In previous publications \cite{nonlinear features BB,unexpected} we have
presented and discussed extensively the unusual nonlinear dynamics observed in
our nonlinear NbN superconducting resonators, which include among others
abrupt bifurcations in the resonance lineshape, hysteresis loops changing
direction, magnetic field sensitivity, resonance frequency shift and also
nonlinear coupling \cite{nonlinear coupling}. These nonlinear dynamics, as it
was shown in \cite{unexpected}, are likely to originate from weak links
forming at the boundaries of the NbN columnar structure. In the present work,
we examine these nonlinear resonators from another aspect using the
intermodulation measurement, which is considered one of the effective tools
for detecting and studying nonlinearities in microwave superconducting devices
in general [4-12], though an intermodulation theory based on extrinsic origins
in superconductors is still lacking \cite{origin impedance,nb}.

The results of intermodulation measurements of these resonators not only
provide an important insight as to the possible nonlinear mechanisms
responsible for the observed dynamics \cite{im theory,physical}, for example,
by providing information about the time scales characterizing the device
nonlinearity \cite{unexpected}, they also exhibit interesting unique features
as well. We show that driving the nonlinear resonator to its onset of
instability while injecting two closely spaced unequal tones laying within the
resonance band into the resonator, results in high amplification of both, the
low amplitude injected signal and the idler (the signal generated via the
nonlinear frequency mixing of the resonator). In Ref. \cite{Yurke}, where the
case of an intermodulation amplifier based on nonlinear Duffing oscillator has
been analyzed, it was shown that intermodulation divergence is expected as the
oscillator is driven near critical slowing down point, where the slope of the
device response with respect to frequency becomes infinite. The fact that our
NbN resonators do not exhibit Duffing oscillator nonlinearity of the kind
employed in the analysis of Ref. \cite{Yurke}, but yet show high
intermodulation gains in the vicinity of the bifurcation points, highly
suggests that the intermodulation gain effect predicted in \cite{Yurke} is not
unique for the Duffing oscillator, and can be demonstrated using other kinds
of nonlinear bifurcations. Moreover in recent publications by Siddiqi
\textit{et al.} \cite{Direct,rf driven}, where dynamical bifurcations between
two driven oscillation states of a Josephson Junction have been directly
observed for the first time, it has been suggested to employ this Josephson
junction nonlinear mechanism for the purpose of amplification and quantum
measurements \cite{rf driven}. \
\begin{figure}
[ptb]
\begin{center}
\includegraphics[
height=0.9755in,
width=2.0081in
]%
{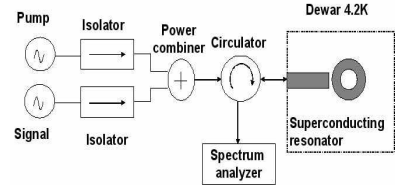}%
\caption{A schematic plot of the intermodulation setup used. }%
\label{imsetup}%
\end{center}
\end{figure}

\begin{figure}
[ptb]
\begin{center}
\includegraphics[
height=2.5659in,
width=3.4272in
]%
{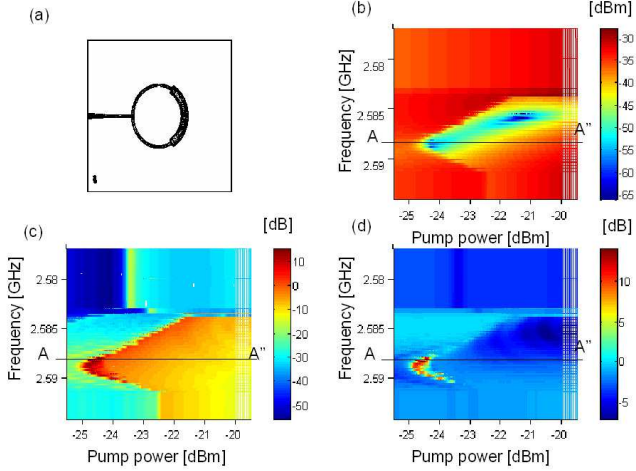}%
\caption{Intermodulation measurement results of B1 first mode. Plot (a)
depicts the layout of B1 resonator. Plots (b), (c) and (d) exhibit meshes of
reflected pump power, idler gain and signal gain respectively, as a function
of transmitted pump power and pump frequency. The meshes were obtained while
gradually decreasing the pump power. The frequency offset between the pump and
signal was set to $2\mathrm{kHz},$ whereas the signal power was set to
$-60$ dBm. The cross sections A-A" are shown in Fig. \ref{crosssection6}.}%
\label{psi6}%
\end{center}
\end{figure}

\begin{figure}
[ptb]
\begin{center}
\includegraphics[
height=1.6708in,
width=2.2252in
]%
{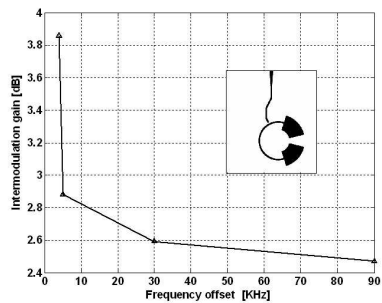}%
\caption{Maximum intermodulation gain achieved during intermodulation
operation as a function of frequency offset set between the pump and signal.
The resonance frequency employed in this measurement was the second mode of B2
resonator. The frequency offsets applied were $4\mathrm{kHz}$,
$5\mathrm{kHz}$, $30\mathrm{kHz}$, $90\mathrm{kHz}$, whereas
the signal power was set to $-60$ dBm. The solid line connecting the data
points is a guideline only. The inset shows the layout of B2 resonator.}%
\label{freqoff}%
\end{center}
\end{figure}

The intermodulation measurements presented in this paper were performed on two
nonlinear NbN superconducting stripline microwave resonators. The layouts of
these resonators, which we will refer to them by the names B1 ($T_{c}=10.7%
\mathrm{K}%
$) and B2 ($T_{c}=6.8%
\mathrm{K}%
$) for simplicity, are depicted in Fig. \ref{psi6} (a), and in the inset of
Fig. \ref{freqoff} respectively. The NbN resonator films were dc-magnetron
sputtered on $34%
\mathrm{mm}%
$X $30%
\mathrm{mm}%
$X $1%
\mathrm{mm}%
$ sapphire substrates near room temperature. The thickness of B1 and B2
resonators are $2200%
\mathrm{\text{\AA}}%
$ and $3000%
\mathrm{\text{\AA}}%
$ respectively. The films were patterned using standard photolithography
process and etched by Ar ion-milling. The coupling gap between B1 and B2
resonators and their feedline was set to $0.4$ $%
\mathrm{mm}%
$ and $0.5$ $%
\mathrm{mm}%
$ respectively. The fabrication process parameters as well as other design
considerations can be found elsewhere \cite{nonlinear features BB}.

The basic intermodulation experimental setup that has been used, is
schematically depicted in Fig. \ref{imsetup}. The input field of the resonator
is composed of two sinusoidal fields generated by external microwave
synthesizers and combined using power combiner. The isolators in the signal
paths were added to minimize crosstalk noise between the signals and suppress
reflections. The signals used have unequal amplitudes, one, which we will
refer to as the pump, is an intense sinusoidal field with frequency $f_{p},$
whereas the other, which we will refer to as the signal, is a small amplitude
sinusoidal field with frequency $f_{p}$ $+f,$ where $f$ represents the
frequency offset between the two signals. Due to the nonlinearity of the
resonator, frequency mixing between the pump and the signal yields an output
idler field at frequency $f_{p}$ $-f.$ Thus the output field from the
resonator, which is redirected by a circulator and measured by a spectrum
analyzer, consists mainly of three spectral components, the reflected pump,
the reflected signal and the generated idler. The intermodulation
amplification in the signal and idler is obtained, as it is shown in this
paper, by driving the resonator to its onset of instability, via tuning the
pump power.%

%

\begin{figure}
[ptb]
\begin{center}
\includegraphics[
height=2.3385in,
width=3.4117in
]%
{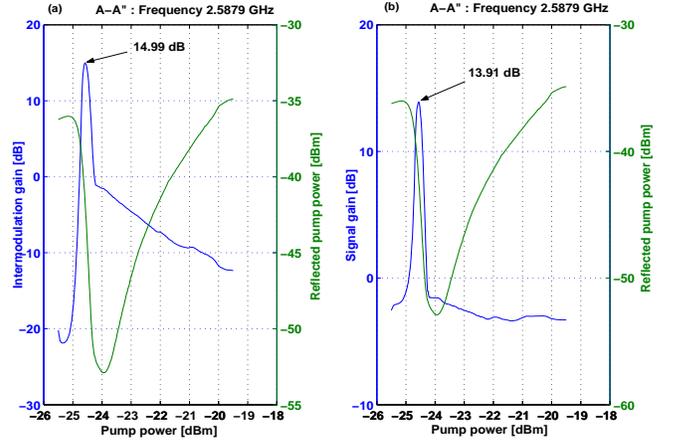}%
\caption{The idler and signal gains at the A-A" cross section of Fig.
\ref{psi6} are shown in plot (a) and (b) respectively, as a function pump
power. The reflected pump power at the same cross section A-A" is also drawn
on the same axis for comparison. A simultaneous amplification in the idler and
signal is measured at the onset of instability of the reflected pump power.}%
\label{crosssection6}%
\end{center}
\end{figure}

In the intermodulation measurements, we limit the signal power to be several
orders of magnitude smaller than the pump power as was assumed in
\cite{Yurke}, and require that all of the tones (pump, signal, idler) lie
within the resonance band of the resonator during the intermodulation operation.

In Fig. \ref{psi6} we present an intermodulation measurement applied to the
first resonance mode of B1 resonator ( $\sim$ $2.58%
\mathrm{GHz}%
$)$,$ at $4.2%
\mathrm{K}%
,$ where we measured the idler and the reflected signals (pump and signal) as
a function of both the transmitted pump power and frequency. The experimental
results presented here were obtained while decreasing the pump power gradually
at each given frequency. The pump power range was set to include the onset of
nonlinear bifurcations of B1 first mode which occurs at relatively low input
powers of the order of $-25$ dBm, whereas the signal was set to a constant
power level of $-60$ dBm, several orders of magnitude lower than the pump. The
pump-signal frequency offset $f$ was set to $2%
\mathrm{kHz}%
,$ very much narrower than the resonance band, (thus ensuring that all three
signals lie within the resonance lineshape during the measurement process).%

\begin{figure}
[ptb]
\begin{center}
\includegraphics[
height=2.6039in,
width=3.3901in
]%
{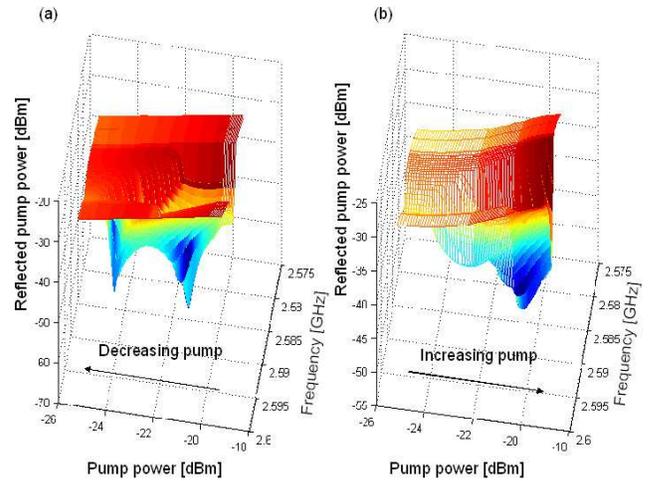}%
\caption{Reflected pump hysteresis during intermodulation operation. Left mesh
obtained while decreasing pump power, and right mesh while increasing pump
power.}%
\label{powerhys6}%
\end{center}
\end{figure}

The reason for varying the pump power rather than its frequency to the edge of
bifurcation, is mostly because the bifurcations along the frequency axis are
abrupt \cite{nonlinear features BB} in contrast to the bifurcations along the
power axis which are more gradual. In Fig. \ref{psi6}, plots (b), (c) and (d)
show the reflected pump power, idler gain and signal gain respectively, as
colormaps of pump power and pump frequency. Large amplifications of the idler
and the signal are measured simultaneously as the reflected pump power
bifurcates, at a given pump frequency, as its power is decreased below some
power threshold. These amplification peaks can be better seen in Fig.
\ref{crosssection6}, where plots (a) and (b) show the A-A" cross-sections of
the idler gain and the signal gain meshes respectively, plotted on the same
axis with the reflected pump power at $2.5879%
\mathrm{GHz}%
$ for comparison.

The amplification gain [dB], which is defined as the difference between the
idler or signal power at the resonator output [dBm] and the signal power [dBm]
at the resonator input (losses in cables and passive devices are calibrated),
reaches $14.99$ dB at its peak in the case of the idler gain, and $13.91$ dB
in the case of the signal gain.

In Fig. \ref{powerhys6} we show an unexpected power-frequency hysteresis of
the reflected pump signal, which implies that the nonlinear resonance shape of
the resonator, as a two dimensional function of input power and frequency, is
multi valued, therefore a care must be taken in choosing the path in reaching
each point in the power-frequency plane. Furthermore in the forward sweep of
the pump no positive gain has been detected in the idler or signal, this may
be partly due to the less steep slopes associated with bifurcations in the
forward direction as seen in Fig. \ref{powerhys6} (b).

Based on these high gains demonstrated experimentally at $T=4.2$ $%
\mathrm{K}%
$ (and $f\sim$ $2.5%
\mathrm{GHz}%
),\ $it is interesting to consider the feasibility of demonstrating some
important quantum phenomena using these nonlinear effects in the quantum
regime where$\ \hbar\omega\gg k_{B}T$ ($T\ll100$ m$%
\mathrm{K}%
$). As in Ref. \cite{Yurke}, consider the mode of operation where a homodyne
detection scheme with a local oscillator having the frequency of the pump is
employed to measure the resonator output. The noise floor of the device is
characterized by the power spectrum $P$ of the homodyne detector output, where
the only externally applied input is the pump. In the nonlinear regime of
operation noise squeezing occurs, namely, $P$ becomes dependent periodically
on the phase of the local oscillator $\phi_{LO}$ relative to the phase of the
pump. In particular, for an intermodulation amplifier having a gain larger
than unity, as the ones described in the present work, the maximum value of
$P\left(  \phi_{LO}\right)  $ may become larger than the value corresponding
to equilibrium noise. In the quantum limit, where $\hbar\omega\gg k_{B}T$,
this effect is somewhat similar to the well known dynamical Casimir effect
\cite{Dodonov}, where a parametric excitation is employed to amplify vacuum
fluctuations and to create real photons.%

According to the idler/signal gain analysis carried out for the case of
Duffing oscillator in Ref. \cite{Yurke}, the idler/signal gain is expected to
diverge as the oscillator is driven to its critical point in the limit
$f\rightarrow0$ . In order to examine whether similar gain-frequency offset
relation holds in our nonlinear resonators, we employed the second resonance
mode of B2 resonator depicted in the inset of Fig. \ref{freqoff}, (which
showed similar behavior under intermodulation operation as B1 resonator,
though with lower gain levels). We varied the frequency offset applied between
the two signals and recorded the maximum idler gain measured at that offset.
The results in Fig. \ref{freqoff}, show clearly that the maximum idler gain
tends to increase as the frequency offset $f$ decreases, though the change was
generally moderate compared with the $1/f$ dependence
predicted in \cite{Yurke}.%

\begin{figure}
[ptb]
\begin{center}
\includegraphics[
height=2.3531in,
width=3.1756in
]%
{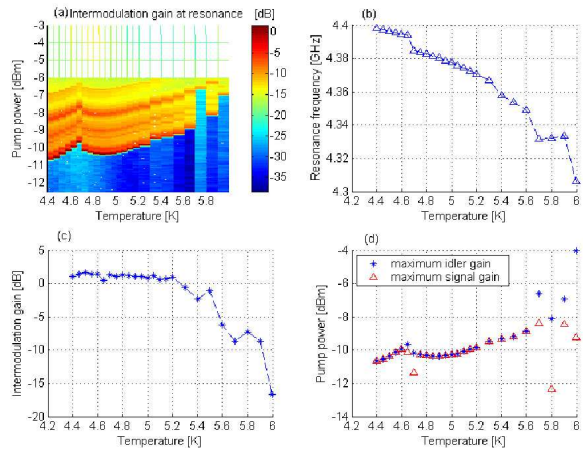}%
\caption{Intermodulation dependence on temperature. Plot (a) exhibits idler
gain as a function of temperature and pump power. Plot (b) shows the shift of
the resonance frequency towards lower frequencies as the temperature is
increased. Plot (c) shows the maximum idler gain measured at different
temperatures. The idler gain is almost constant at low temperatures. Plot (d)
exhibits the good correlation that exists between the idler and signal maximum
gain coordinations in the temperature-pump power plane below
$5.6\mathrm{K}.$ }%
\label{tempanaly}%
\end{center}
\end{figure}

Moreover we studied the dependence of intermodulation gain on temperature
using B2 resonator and its second mode. We increased the ambient temperature
in small steps between $4.4%
\mathrm{K}%
$ and $6%
\mathrm{K}%
$ (below $T_{c}=6.8%
\mathrm{K}%
$)$.$ At each given temperature we set the pump frequency to the resonance
frequency of the resonator corresponding to that temperature, and measured the
idler and signal gains while gradually decreasing the pump power. The
frequency offset between the pump and signal was set to $120%
\mathrm{kHz}%
$, whereas the signal power was set to $-55$ dBm. Fig. \ref{tempanaly}
summarizes the main results of this measurement. In plot (a) the idler gain
mesh is shown as a function of temperature and pump power. The idler maximum
gain is generally obtained (earlier) at higher pump powers as the temperature
increases. In plot (b) the resonance frequency of the resonator is depicted as
a function of temperature. The resonance frequency shifts towards lower
frequencies as the temperature increases, due to increase in the nonlinear
inductance and the penetration depth of the resonator as the temperature is
increased \cite{nb,hairpin temp}. In plot (c) the maximum measured idler gain
is plotted as a function of temperature. A plateau in the maximum idler gain
is observed for low temperatures, but it decreases considerably as the
temperature exceeds $5.2%
\mathrm{K}%
$. Such plateau in measured intermodulation powers at low and intermediate
temperatures have been reported in \cite{hairpin temp,Koren}, nevertheless
whereas in \cite{hairpin temp,Koren,im theory} the intermodulation power
increases due to decrease in the intermodulation critical current density
$J_{IMD}$ or increase in the kinetic inductance as the temperature is
increased, our results show a decrease in the measured intermodulation gain as
a function of temperature. This discrepancy can be attributed to the observed
degradation of the device response (slope), responsible for the amplification,
due to increase in the film surface resistance as the temperature is
increased. In plot (d) the coordinations of the maximum measured idler and
signal gains in the temperature-pump power plane are shown. The blue stars
represent maximum idler gain coordination whereas the red triangles represent
maximum signal gain coordination. The plot exhibit an excellent correlation
between the two coordinations up to about $5.6%
\mathrm{K}%
.$

In conclusion, we have measured intermodulation gain in two of our nonlinear
NbN superconducting stripline resonators, at relatively low temperatures
$\sim$ $4.2%
\mathrm{K}%
.$ An intermodulation gain as high as $\sim$ $15$ dB, was successfully
achieved in one of the resonators, while applying $2%
\mathrm{kHz}%
$ frequency offset and $-60$ dBm signal power. Moreover we showed that the
reflected pump power, the signal gain and the idler gain demonstrate strong
hysteretic behavior in the frequency-pump power plane. In addition, we briefly
discussed the dependence of the intermodulation gain on ambient temperature,
and frequency offset applied between the pump and signal. The intermodulation
gain results were found to be both reproducible and controllable, which is a
preliminary condition for any practical application. Whereas the underlying
physics remains an outstanding challenge for future research, these nonlinear
resonators operated as intermodulation amplifiers may be potentially employed,
in the future, in generating low noise microwave signals, signal switching,
and even in, producing quantum squeezed states \cite{Yurke} and amplifying
quantum zero point fluctuations.

E.B. would especially like to thank Michael L. Roukes for supporting the early
stage of this research and for many helpful conversations and invaluable
suggestions. Very helpful conversations with Bernard Yurke are also gratefully
acknowledged. This work was supported by the German Israel Foundation under
grant 1-2038.1114.07, the Israel Science Foundation under grant 1380021, the
Deborah Foundation and Poznanski Foundation.

\bibliographystyle{plain}
\bibliography{apssamp}

\end{document}